\documentstyle[aps,prd,twocolumn,axodraw,epsfig]{revtex}
                                                                                                          
\def\beq{\begin{equation}}
\def\eeq#1{\label{#1}\end{equation}}
\def\eeqn{\end{equation}}
\def\beqa{\begin{eqnarray}}
\def\eeqa#1{\label{#1}\end{eqnarray}}
\def\eeqan{\end{eqnarray}}
\def\CR{\nonumber \\ }
\def\leqn#1{(\ref{#1})}
\def\({\left(}
\def\){\right)}

\def\stacksymbols #1#2#3#4{\def\theguybelow{#2}
    \def\vp{\lower#3pt}
    \def\sp{\baselineskip0pt\lineskip#4pt}
    \mathrel{\mathpalette\intermediary#1}}

\def\intermediary#1#2{\vp\vbox{\sp
     \everycr={}\tabskip0pt
     \halign{$\mathsurround0pt#1\hfil##\hfil$\crcr#2\crcr
              \theguybelow\crcr}}}

\def\lapproxeq{\stacksymbols{<}{\sim}{2.5}{.2}}

\def\O{{\cal O}}

\def\Mz{M_{\rm Z}}
\def\Mw{M_{\rm W}}

\newcommand{\bspace}{\!\!\!\!}
\newcommand{\met}{\mbox{$E{\bspace}/_{T}$}}

\def\mini#1{\leavevmode\hbox{\tiny #1}}

\def\g#1{g_{\mini{#1}}}

\def\to{\rightarrow}

\begin{document}

\wideabs{

\title{Collider Phenomenology of the Higgsless Models}
\author{Andreas Birkedal$^{1}$, Konstantin Matchev$^{1}$, and Maxim Perelstein$^2$}
\address{$^1$ Physics Department, University of Florida, Gainesville, FL 32611\\
$^2$ Institute for High-Energy Phenomenology, Cornell University, Ithaca, NY~14853}

\date{April 5, 2005}
\maketitle

\begin{abstract}
We identify and study the signatures of the recently proposed Higgsless models 
at the Large Hadron Collider (LHC). We concentrate on tests of 
the mechanism of partial unitarity restoration in the longitudinal 
vector boson 
scattering, which is crucial to the phenomenological success of any Higgsless 
model and does not depend on the model-building details. We investigate the
discovery reach for charged massive vector boson resonances and show that
all of the preferred parameter space will be probed with $100\ {\rm fb}^{-1}$
of LHC data. Unitarity restoration requires that the masses and couplings of 
the resonances obey certain sum rules. We discuss the prospects for their
experimental verification at the LHC.
\end{abstract}
}

{\em Introduction ---} Numerous experiments in high-energy physics confirm 
that the electromagnetic and weak interactions can be understood in terms 
of a non-abelian gauge theory with spontaneously broken "electroweak" 
$SU(2)_L\times U(1)_Y$ symmetry. The mechanism of the electroweak 
symmetry breaking (EWSB) is at present unknown. The suggested 
mechanisms may be roughly divided into those involving only weakly 
coupled physics (of which the Standard Model, perhaps supplemented 
by supersymmetry, is the best-known example), and those that rely 
on strong dynamics to break the symmetry, as in technicolor models~\cite{TC}. 
A simple estimate for the scale of strong dynamics in theories 
of the second type, based on the unitarity violation in the 
scattering of longitudinal massive gauge bosons, gives 
\beq
\Lambda\sim4\pi M_W/g \sim 1.8~{\rm TeV},
\eeq{lambda}
where $g$ is the $SU(2)_L$ gauge coupling. The strongly coupled physics 
at this scale is generically ruled out by precision electroweak 
constraints (PEC), seemingly disfavoring the idea of the EWSB by 
strong dynamics~\cite{PT}. Recently, however, this idea was implemented in 
a novel way in the "Higgsless" models~\cite{KK1,KK2,Nomura,KK3}. 
In these models, new weakly coupled particles appear at the TeV scale
and postpone unitarity violation~\cite{SekharChivukula:2001hz}, 
raising the strong coupling scale well above the naive 
estimate~\leqn{lambda}. As a result, the models have a better chance 
of achieving consistency with the PEC. In fact, while the original 
Higgsless models did not allow to raise $\Lambda$ significantly 
without running into conflict with PEC~\cite{KKisnotOK,DHLR,PEWothers}, 
it was recently shown that $\Lambda$ can be raised by at least a factor 
of 10 with appropriate modifications of the fermion sector~\cite{KKisOK}. 

In this letter, we would like to identify and study the collider signatures 
of the Higgsless models at the Large Hadron Collider (LHC). In realistic models, 
the strong coupling scale $\Lambda$ is too high for the LHC to directly observe 
the strongly coupled physics responsible for the EWSB. On the other hand, the 
additional weakly coupled states required to raise $\Lambda$ will be observable. 
A number of Higgsless models have been proposed in the literature, differing 
in the number of spatial dimensions (five in the original models, four in 
the "deconstructed" versions~\cite{decon}), the embedding of the 
Standard Model (SM) fermions, and so on. 
The spectrum of the TeV scale states and their interactions with the 
SM particles are highly model-dependent. The {\em mechanism} by which 
$\Lambda$ is raised, however, is the same in all models: new massive 
vector bosons (MVBs), with the same quantum numbers as the familiar 
$W$, $Z$ and $\gamma$, appear at the TeV scale. The  couplings between 
the MVBs and the SM $W/Z$ gauge bosons obey (at least approximately) 
{\it unitarity sum rules}, which enforce the cancellation 
of the fast-growing terms in the longitudinal gauge boson 
scattering amplitudes, postponing the unitarity violation. 
We will concentrate on the experimental signatures that 
test this mechanism, and are therefore independent of the 
details of specific Higgsless models. 

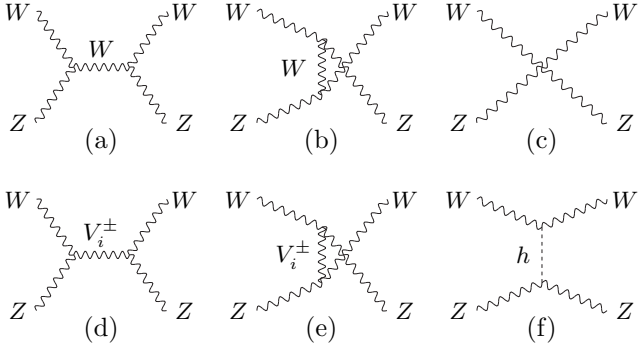
\begin{figure}[t!]
\begin{center} 
{
\unitlength=0.7 pt
\SetScale{0.7}
\SetWidth{0.5}      
\normalsize    
{} \allowbreak
\begin{picture}(100,100)(0,0)
\Photon(15,80)(35,50){2}{6}
\Photon(15,20)(35,50){2}{6}
\Photon(35,50)(65,50){2}{6}
\Photon(65,50)(85,80){2}{6}
\Photon(65,50)(85,20){2}{6}
\Text(50,10)[c]{(a)}
\Text(50,60)[c]{\small $W$}
\Text( 5,80)[c]{\small $W$}
\Text( 5,20)[c]{\small $Z$}
\Text(95,80)[c]{\small $W$}
\Text(95,20)[c]{\small $Z$}
\end{picture}
\quad
%
\begin{picture}(100,100)(0,0)
\Photon(15,80)(50,65){2}{6}
\Photon(15,20)(50,35){2}{6}
\Photon(50,65)(50,35){2}{6}
\Photon(50,65)(85,20){2}{9}
\Photon(50,35)(85,80){2}{9}
\Text(50,10)[c]{(b)}
\Text(35,50)[c]{\small $W$}
\Text( 5,80)[c]{\small $W$}
\Text( 5,20)[c]{\small $Z$}
\Text(95,80)[c]{\small $W$}
\Text(95,20)[c]{\small $Z$}
\end{picture}\
\quad
%
\begin{picture}(100,100)(0,0)
\Photon(15,80)(85,20){2}{13}
\Photon(15,20)(85,80){2}{13}
\Text(50,10)[c]{(c)}
\Text( 5,80)[c]{\small $W$}
\Text( 5,20)[c]{\small $Z$}
\Text(95,80)[c]{\small $W$}
\Text(95,20)[c]{\small $Z$}
\end{picture}\
%
\begin{picture}(100,100)(0,0)
\Photon(15,80)(35,50){2}{6}
\Photon(15,20)(35,50){2}{6}
\Photon(35,50)(65,50){2}{6}
\Photon(65,50)(85,80){2}{6}
\Photon(65,50)(85,20){2}{6}
\Text(50,10)[c]{(d)}
\Text(50,62)[c]{\small $V_i^\pm$}
\Text( 5,80)[c]{\small $W$}
\Text( 5,20)[c]{\small $Z$}
\Text(95,80)[c]{\small $W$}
\Text(95,20)[c]{\small $Z$}
\end{picture}\
\quad
%
\begin{picture}(100,100)(0,0)
\Photon(15,80)(50,65){2}{6}
\Photon(15,20)(50,35){2}{6}
\Photon(50,65)(50,35){2}{6}
\Photon(50,65)(85,20){2}{9}
\Photon(50,35)(85,80){2}{9}
\Text(50,10)[c]{(e)}
\Text(35,50)[c]{\small $V_i^\pm$}
\Text( 5,80)[c]{\small $W$}
\Text( 5,20)[c]{\small $Z$}
\Text(95,80)[c]{\small $W$}
\Text(95,20)[c]{\small $Z$}
\end{picture}\
\quad
%
\begin{picture}(100,100)(0,0)
\Photon(15,80)(50,65){2}{6}
\Photon(15,20)(50,35){2}{6}
\DashLine(50,65)(50,35){2}
\Photon(50,65)(85,80){2}{6}
\Photon(50,35)(85,20){2}{6}
\Text(50,10)[c]{(f)}
\Text(40,50)[c]{\small $h$}
\Text( 5,80)[c]{\small $W$}
\Text( 5,20)[c]{\small $Z$}
\Text(95,80)[c]{\small $W$}
\Text(95,20)[c]{\small $Z$}
\end{picture}
}
\end{center}
\caption{Diagrams contributing to the $W^\pm Z\to W^\pm Z$ scattering process: 
(a), (b) and (c) appear both in the SM and in
Higgsless models, (d) and (e) only appear in Higgsless models, while
(f) only appears in the SM.}
\label{diagram}
\end{figure}

{\em Unitarity Sum Rules---} Consider the elastic scattering 
process $W^\pm_L Z_L\to W^\pm_L Z_L$. 
In the absence of the Higgs boson, this process receives contributions 
from the three Feynman diagrams shown in Figs.~\ref{diagram} (a) -- (c). 
Evaluating these diagrams yields the amplitude
\beqa
&&{\cal M}^{\pm0\pm0} = (\g{WWZZ} - \g{WWZ}^2)\cdot\Bigl[(c^2-6c-3)E^4 \nonumber \\
&&+\, (c^2-3c-2)\Mz^2 E^2\,-\,(c^2-9c-4)\Mw^2 E^2 \Bigr] \nonumber \\
&&+\, \g{WWZ}^2 \frac{\Mz^4(1-c)}{2\Mw^2} E^2 \,+\,\O(E^0), 
\eeqa{WZWZsansH}
where $E\gg M_{\rm W,Z}$ is the energy of the incoming $W$ boson 
in the center of mass frame, $c$ is the cosine of the scattering angle 
(the angle between the incoming and outgoing $W$ bosons), the overall 
factor of $i\Mw^{-2}\Mz^{-2}$ has been omitted, and the notation used 
for the coupling strengths is self-explanatory. In the Higgsless 
theories, this process receives an additional contribution from 
the diagrams in Figs.~\ref{diagram} (d) and (e), where $V_i^\pm$ denotes the 
charged MVB of mass $M_i^\pm$ (the Lorentz structure of the $V^\pm W^\mp Z$ 
coupling is identical to the familiar SM $W^\pm W^\mp Z$ vertex). 
The index $i$ corresponds to the KK 
level of the state in the case of a 5D theory, or labels the 
mass eigenstates in the case of a 4D deconstructed theory. 
At high energies ($E\gg M_i^\pm$) the contribution of $V_i^\pm$ is given by
\beqa
&&\Delta {\cal M}^{\pm0\pm0}_{\rm V} = -(\g{WZV}^{(i)})^2 
\Bigl[(c^2-6c-3)E^4 \nonumber \\
&& + (c^2-2c-3)\Mz^2 E^2 - (c^2-10c-3)\Mw^2 E^2  \nonumber \\
&& +(1-c)\ \frac{3\left(M_i^\pm\right)^4-\left(\Mw^2-\Mz^2\right)^2}{2\left(M_i^\pm\right)^2} E^2 \Bigr]+\,\O(E^0).
\eeqa{WZWZV}
The notation for the three-point coupling strength is again 
self-explanatory, and the overall factor of $i\Mw^{-2}\Mz^{-2}$ has been dropped. 
Note that there is no diagram involving neutral MVBs, $V_i^0$ --- 
the quantum numbers forbid three-point and four-point couplings 
involving exclusively neutral states. Remarkably, the $E^4$ and $E^2$
terms in Eq.~\leqn{WZWZsansH} can be exactly cancelled 
by the contribution of the MVBs, provided that the following sum rules are satisfied:
\beqa
& &\g{WWZZ} = \g{WWZ}^2 \,+\, \sum_i (\g{WZV}^{(i)})^2, \CR
&2&(\g{WWZZ}-\g{WWZ}^2)(\Mw^2+\Mz^2) + \g{WWZ}^2\,\frac{\Mz^4}{\Mw^2}\CR 
&=& \sum_i (\g{WZV}^{(i)})^2\,
\left[3(M^\pm_i)^2-\frac{(\Mz^2-\Mw^2)^2}{(M^\pm_i)^2}\right]. 
\eeqa{sumW} 
\noindent
In 5D theories, these equations are satisfied exactly if all the KK states, 
$i=1\ldots\infty$, are taken into account. This is not an accident, but a 
consequence of the gauge symmetry and locality of the underlying theory. 
While this is not sufficient to ensure unitarity at all energies 
(the increasing number of inelastic channels ultimately results 
in unitarity violation), the strong coupling scale can be 
significantly higher than the naive estimate~\leqn{lambda}. 
For example, in the warped-space Higgsless models~\cite{KK2,KKisOK} 
unitarity is violated at the scale~\cite{Pap} 
\beq
\Lambda_{\rm NDA} \sim \frac{3\pi^4}{g^2}\frac{\Mw^2}{M^\pm_1},
\eeq{NDA}
which is typically of order 5--10 TeV. In 4D models, the number of 
the MVBs is finite and the second of the sum rules~\leqn{sumW} is 
only satisfied approximately; however, a numerical study of sample models 
indicates that the violation of the sum rule has to be very small 
(at the level of 1\%) to achieve an adequate improvement in $\Lambda$~\cite{us_long}. 

Considering the $W^+_LW^-_L\to  W^+_LW^-_L$ scattering process yields the sum 
rules constraining the couplings of the neutral MVBs $V_i^0$~\cite{KK1}:
\beqa
\g{WWWW} &=& \g{WWZ}^2 + \g{WW$\gamma$}^2 \,+\, \sum_i (\g{WWV}^{(i)})^2, \CR
4\g{WWWW}\,\Mw^2 &=& 3\,\left[\g{WWZ}^2 \Mz^2 + \sum_i (\g{WWV}^{(i)})^2
\,(M^0_i)^2\right],
\eeqa{sumZ} 
where $M_i^0$ is the mass of the $V_i^0$ boson.
Considering other channels such as
$W_L^+W_L^-\to ZZ$ and $ZZ\to ZZ$ does not yield any new sum rules.
The presence of multiple MVBs, whose couplings obey Eqs.~\leqn{sumW},~\leqn{sumZ}, 
is a generic prediction of the Higgsless models. 

{\em Collider Phenomenology---} Our study of the collider phenomenology 
in the Higgsless models will focus on the vector boson fusion processes. 
These processes are attractive for two reasons. Firstly, the production 
of the MVBs via vector boson fusion is relatively model-independent, since 
the couplings are constrained by the sum rules~\leqn{sumW},~\leqn{sumZ}. 
This is in sharp contrast with the 
Drell-Yan production mechanism~\cite{DHLR}, which dominates for the 
conventional $W^\prime$ and $Z^\prime$ bosons but is likely to be 
suppressed for the Higgsless MVBs due to their small couplings to 
fermions, which is needed to evade PEC~\cite{KKisOK}. 
Secondly, if enough couplings and masses can be measured, 
these processes can provide a {\em test} of the sum rules, probing 
the mechanism of partial unitarity restoration.

Eq.~\leqn{NDA} indicates that the first MVB should appear 
below $\sim 1$ TeV, and thus be accessible at the LHC. 
For $V^\pm_1$, the sum rules~\leqn{sumW} imply an inequality
\beq
\g{WZV}^{(1)}  
\lapproxeq \frac{\g{WWZ}\Mz^2}{\sqrt{3}M^\pm_1\Mw}.
\eeq{bound}
This bound is quite stringent ($\g{WZV}^{(1)}\lapproxeq0.04$ 
for $M^\pm_1=700$ GeV). 
At the same time, convergence of the sum rules in~\leqn{sumW} requires 
\beq
\g{WZV}^{(k)}\,\propto\,k^{-1/2}\,(M^\pm_k)^{-1}.
\eeq{drops}
The combination of heavier masses and lower couplings means that 
the heavier MVBs may well be unobservable, so that only the 
$V_1$ states can be studied. On the other hand, a numerical 
study of sample models indicates that the unitarity sum rules 
converge quite rapidly~\cite{us_long}. The "saturation limit", in which there 
is only a single set of MVBs whose couplings saturate the sum 
rules, is likely to provide a good approximation to the 
phenomenology of the realistic Higgsless models. 
In this limit, the partial width of the $V^\pm_1$ is given by
\begin{equation}
\Gamma(V^\pm_1\to W^\pm Z)\approx \frac{\alpha\ (M^\pm_1)^3}{144\, s_w^2\, M_W^2}\, ,
\label{width}
\end{equation}
where $s_w\equiv \sin\theta_W$ is sine of the Weinberg angle.
This formula is analogous, but not identical, to the 
well-known KSFR relation \cite{KSFR}.

\begin{figure}[tb]
\begin{center}
\includegraphics[scale=0.47]{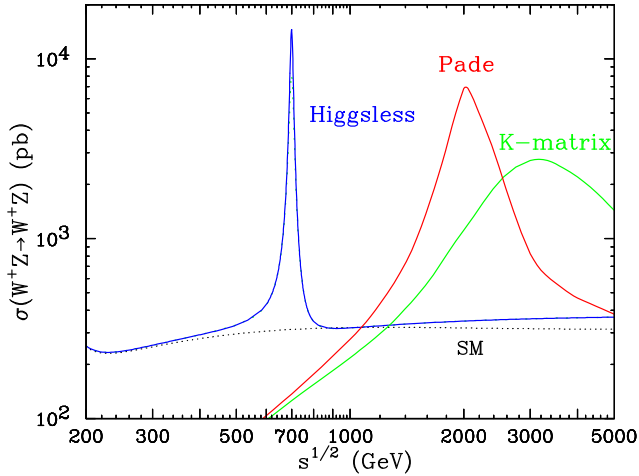}
\vskip2mm
\caption{$WZ$ elastic scattering cross-sections in the SM (dotted), 
the Higgsless model (blue), and two "unitarization" models: 
Pad\'e (red) and K-matrix (green).}
\label{fig:WZ}
\end{center}
\end{figure}

A particularly interesting scattering channel is $WZ\to WZ$. 
In this channel, the Higgsless model predicts a series of 
resonances, see Fig.~\ref{diagram}(d). In the Standard Model, 
on the other hand, this amplitude is unitarized by the $t$-channel 
Higgs exchange as in Fig.~\ref{diagram}(f), and has no resonance. 
Conventional theories of 
EWSB by strong dynamics may contain a resonance in this channel, 
but it is likely to be heavy ($\sim 2$ TeV for QCD-like theories) 
and broad due to strong coupling.  
An illustration is provided by Fig.~\ref{fig:WZ}, showing 
the parton-level cross section for this process in a 
Higgsless model in the saturation limit with a 700 GeV charged MVB $V^\pm_1$. 
We assume that the $V_1^\pm$ has no significant couplings to fermions.
With these assumptions the $V_1^\pm$ width is about 15 GeV.
For comparison, the figure shows the cross section in the SM with 
a $700$ GeV Higgs, and in two phenomenological "unitarization models" 
which attempt to mimic the physics of the conventional 
technicolor-type theories: the Pad\'e approximant and 
K-matrix schemes defined in~Ref.\cite{DH} and available in the {\tt PYTHIA}
general purpose event generator. 
(The parameters used in Fig.~\ref{fig:WZ} were 
obtained in~\cite{DH} by "scaling up" the parameters 
of the pion chiral lagrangian; in the notation of~\cite{DH}, 
$M_R(\mu=2{\rm~TeV})=-9\times 10^{-4}, N_R(\mu=2{\rm~TeV})=1.8\times 10^{-3}.$)

A striking feature of the charged MVB resonance is its small width: the
resonance is almost a factor of 20 narrower than a SM Higgs of the same
mass. This is primarily due to the vector nature of the MVB and the
fact that it only has a single decay channel. At the same time, the
coupling between the MVB and the SM gauge bosons is of a strength
similar to the SM Higgs, as required for the unitarization of the
$I=0,J=0$ channel in the Higgsless models.

\begin{figure}[t!]
\begin{center}
\includegraphics[scale=0.47]{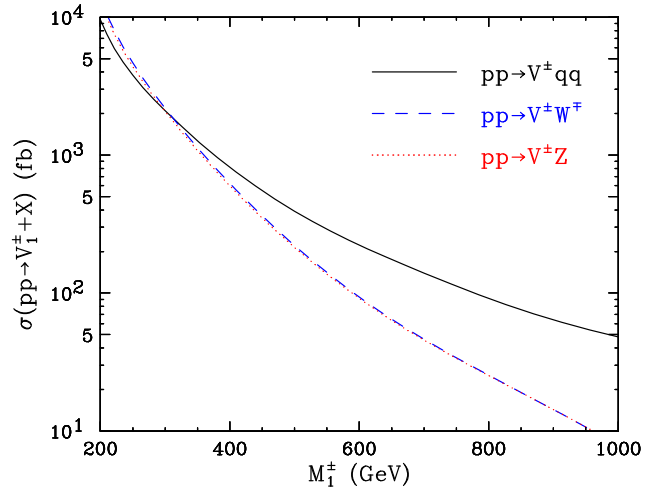}
\vskip2mm
\caption{Production cross-sections of $V^\pm$ at the LHC.}
\label{fig:xsec}
\end{center}
\end{figure}

\begin{figure}[tb]
\begin{center}
\includegraphics[scale=0.47]{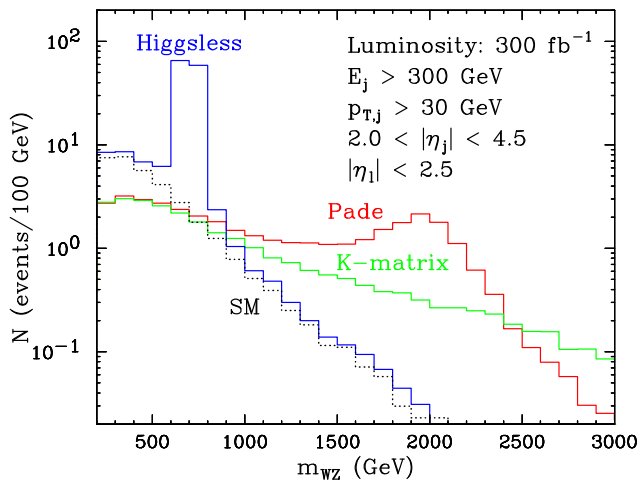}
\vskip2mm
\caption{The number of events per 100 GeV bin in the $2j+ 3\ell+\nu$ 
channel at the LHC with an integrated luminosity of 300
fb$^{-1}$ and cuts as indicated in the figure. The model 
assumptions and parameter choices are the same as in Fig.~\ref{fig:WZ}.}
\label{fig:histo}
\end{center}
\end{figure}

At the LHC, the vector boson fusion processes will occur as a result 
of $W/Z$ bremsstrahlung off quarks. The typical final state for such 
events includes two forward jets in addition to a pair of gauge bosons. 
The production cross section of $V^\pm_1$ in association with two jets
is shown by the solid line in Fig.~\ref{fig:xsec}.
To estimate the prospects for the charged MVB search at the LHC, we 
require that both jets be observable (we assume jet rapidity coverage 
of $|\eta|\leq 4.5$), and impose the following lower cuts on 
the jet rapidity, energy, and transverse momentum: 
$|\eta|>2, E>300$ GeV, $p_T>30$ GeV. These requirements 
enhance the contribution of the vector boson fusion
diagrams relative to the irreducible background of the non-fusion 
$q\bar{q}'\to WZ$ SM process as well as $q\bar{q}'\to V_1^\pm$ Drell-Yan
process. The ``gold-plated'' final state~\cite{Bagger:1993zf}
for this search is $2j+3\ell$+\met, with the 
additional kinematic requirement that two of the leptons have to be consistent 
with a $Z$ decay~\cite{others}. We assume lepton rapidity coverage of
$|\eta|<2.5$. The $WZ$ invariant mass, $m_{WZ}$, 
can be reconstructed using the missing transverse energy measurement and 
requiring that the neutrino and the odd lepton form a $W$. 
The number of "gold-plated" events (including all lepton sign combinations)
in a 300 fb$^{-1}$ LHC data sample, 
as a function of $m_{WZ}$, is shown in Fig.~\ref{fig:histo}. 
For comparison, this figure also shows the predictions of the four 
models discussed above, with the same assumptions as in Fig.~\ref{fig:WZ}. 
The Higgsless model can be easily identified by observing the 
MVB resonance: for the chosen parameters, the dataset contains 
$130$ $V_1^\pm\to W^\pm Z \to 3\ell+\nu$ events. The irreducible 
non-fusion SM background is effectively suppressed by the cuts: 
the entire dataset shown in Fig.~\ref{fig:histo} contains only 
$6$ such events. We therefore estimate the discovery reach 
for $V^\pm_1$ resonance by requiring 10 signal events after cuts.
The efficiency of the cuts for $500\ {\rm GeV}\le M^\pm_1\le 3\ {\rm TeV}$
is in the range $20-25\%$.
We then find that with $10\ {\rm fb}^{-1}$ of data, corresponding 
to 1 year of running at low luminosity, the LHC will probe the Higgsless models 
up to $M^{\pm}_1\lapproxeq 550$ GeV, while covering the whole 
preferred range up to $M^{\pm}_1=1$ TeV requires $60\ {\rm fb}^{-1}$.
Note, however, that one should expect a certain amount of 
reducible background with fake and/or non-isolated leptons.

Once the $V^\pm_1$ resonance is discovered, identifying it as part of 
a Higgsless model requires testing the sum rules (\ref{sumW}) by
measuring its mass $M^\pm_1$ and coupling $g^{(1)}_{WZV}$. 
The coupling can be determined from the total $V^\pm_1$
production cross section $\sigma_{\rm tot}$.
However, we are observing the $V^\pm_1$ resonance in an exclusive channel,
which only yields the product $\sigma_{\rm tot}\,BR(V^\pm_1\to W^\pm Z)$.
A measurement of the total resonance width $\Gamma(V^\pm_1\to{\rm anything})$
would remove the dependence on the unknown branching fraction $BR$.
However, the accuracy of this measurement is severely limited by the
poor missing energy resolution. Nevertheless, a Higgsless origin of the
resonance can be ruled out if the value of $g^{(1)}_{WZV}$,
inferred with the assumption of $BR=1$, violates the bound (\ref{bound}). 

By transferring a $Z$ or a $W^\pm$ from the initial state to the final state in 
Figs.~\ref{diagram} (d) and (e), we obtain an alternative $V^\pm_1$
production process, the associated production, which can be used for discovery 
as well as testing the sum rules (\ref{sumW}).
The total cross section for this process
is shown in Fig.~\ref{fig:xsec}. The $W^\pm ZZ$ 
final state, with the requirement that all three gauge bosons decay leptonically,
yields a very clean $5\ell+\met$ signature. One can then
reconstruct the two $Z$'s and the $V^\pm_1$ resonance. The main advantage of
this purely leptonic channel would be the superior measurement of the total 
width; however, the analysis is statistics limited and the discovery reach 
does not extend beyond $500$ GeV, even for the high-luminosity LHC option.

{\it Conclusions ---} It has long been known that the vector boson fusion processes 
will provide an important tool for testing the strongly coupled theories of EWSB at 
the LHC. This is as true for the recently proposed Higgsless models as it is for 
traditional technicolor theories. As we discussed in this letter, the observation of a 
light and narrow resonance in the $WZ$ channel would be a smoking gun for
the Higgsless models. In addition, the Higgsless models provide 
a robust, definite prediction concerning the 
properties of the resonance, the sum rules (\ref{sumW}), which can also be 
tested in this channel.  

While we have concentrated on the $WZ$ channel which provides the most striking 
signals, other vector boson fusion processes may also be useful. 
The neutral MVBs $V_i^0$ would appear as resonances in the $W^+W^-$ channel;
however, reconstructing these resonances requires 
hadronic $W$ decays and suffers from severe backgrounds~\cite{Butterworth:2002tt}.
The $ZZ$ channel exhibits no resonance, but could provide an 
independent test of the model. These channels will be explored in more detail 
in~\cite{us_long}.


{\it Acknowledgments ---} We would like to thank C.~Cs\'aki and 
G.~Cacciapaglia for helpful discussions. MP is supported by 
NSF grant PHY-0355005. KM and AB are supported by a US DoE 
Outstanding Junior Investigator award under grant DE-FG02-97ER41029. 
MP would like to thank the theory group at the University of Florida 
for their hospitality during the completion of this work.


\begin{thebibliography}{99}
\bibitem{others}
Other channels, such as 
$4j+2l$, may also be useful, although the backgrounds are expected to be 
higher. 
\end{thebibliography}
\end{document}